# Kinematic Flexibility Analysis: Hydrogen Bonding Patterns Impart a Spatial Hierarchy of Protein Motion

Dominik Budday,*,† Sigrid Leyendecker,† and Henry van den Bedem*,‡

†*Chair of Applied Dynamics, University of Erlangen-Nuremberg, Erlangen, Germany*
‡*Biosciences Division, SLAC National Accelerator Laboratory, Stanford University, California, Menlo Park, USA*

E-mail: dominik.budday@fau.de; vdbedem@stanford.edu

## Abstract

Elastic network models (ENM) and constraint-based, topological rigidity analysis are two distinct, coarse-grained approaches to study conformational flexibility of macromolecules. In the two decades since their introduction, both have contributed significantly to insights into protein molecular mechanisms and function. However, despite a shared purpose of these approaches, the topological nature of rigidity analysis, and thereby the absence of motion modes, has impeded a direct comparison. Here, we present an alternative, kinematic approach to rigidity analysis, which circumvents these drawbacks. We introduce a novel protein hydrogen bond network spectral decomposition, which provides an orthonormal basis for collective motions modulated by non-covalent interactions, analogous to the eigenspectrum of normal modes, and decomposes proteins into rigid clusters identical to those from topological rigidity. Our kinematic flexibility analysis bridges topological rigidity theory and ENM, and enables a detailed analysis of motion modes obtained from both approaches. Our analysis reveals that collectivity of protein motions, reported by the Shannon entropy, is significantly lower for rigidity theory versus normal mode approaches. Strikingly, kinematic flexibility analysis suggests that the hydrogen bonding network encodes a protein-fold specific, spatial hierarchy of motions, which goes nearly undetected in ENM. This hierarchy reveals distinct motion regimes that rationalize protein stiffness changes observed from experiment and molecular dynamics simulations. A formal expression for changes in free energy derived from the spectral decomposition indicates that motions across nearly 40% of modes obey enthalpy-entropy compensation. Taken together, our analysis suggests that hydrogen bond networks have evolved to modulate protein structure *and* dynamics.

## Introduction

Coarse-grained modeling techniques can provide significant insight into the dynamic behavior and biological function of macromolecules. Elastic Network Models (ENM[1]) and rigidity theory based flexibility analysis[2,3] are two well known approaches that have been applied extensively to study molecular motion. Although they are based on physically distinct concepts, both aim to distinguish more flexible regions from compact ones in the molecule, and to predict large-scale, functional motions.
ENM (Figure 1, right) approximate the potential energy function $V$ via pairwise harmonic interactions $V = \sum_{ij} C_{ij}(|\mathbf{r}_{ij}| - |\mathbf{r}_{ij}^0|)^2$, where $C_{ij}$ denotes the stiffness[1] or an exponentially fading weight[4] of the restraint between atoms $i, j$

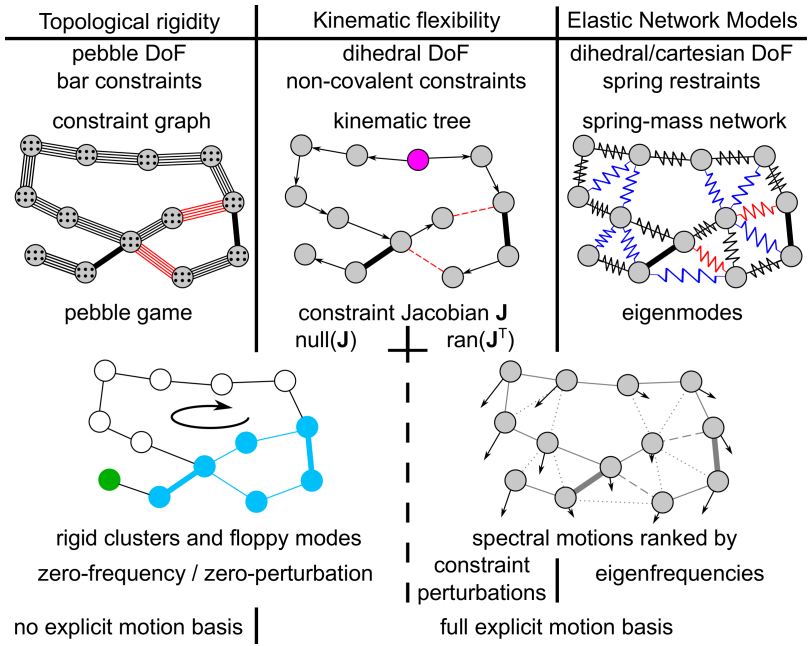

Figure 1: Coarse-grained protein modeling via topological rigidity analysis (left) and elastic network models (ENM, right). Topological rigidity analysis decomposes a protein into rigid clusters of atoms from explicit constraints, quantifying the number of 'floppy modes' without an explicit motion basis. By contrast, ENM obtains an explicit motion basis corresponding to eigenmodes of one-dimensional restraints. Our kinematic flexibility approach (center) combines features from topological rigidity and ENM, providing motion modes from a spectral decomposition of explicit constraints corresponding to, for example, hydrogen bonds.

located at $\mathbf{r}_i, \mathbf{r}_j$, respectively, with $\mathbf{r}_{ij} = \mathbf{r}_j - \mathbf{r}_i$ and rest length $|\mathbf{r}_{ij}^0|$. The resulting spring-mass network can be formulated in terms of dihedral[5,6] or cartesian[1,7–9] degrees of freedom (DoF) and analyzed with classical Hamiltonian mechanics. Diagonalization of the Hessian matrix leads to a spectral distribution of the dynamics in terms of eigenmodes with corresponding eigenfrequencies. The simplified energy function of ENM places the native structure at a global minimum, circumventing initial minimization necessary in traditional normal mode analysis (NMA).[10] The motions corresponding to low-frequency modes are robust, often agree with those of more detailed models,[11,12] and can be functionally relevant.[13] Dihedral-based approaches often correlate better with experimentally observed conformational changes than cartesian-based methods.[14,15] Other variants include Gaussian[16,17] or anisotropic[18] network models.

In rigidity theory, macromolecules are modeled as constraint graphs (Figure 1, left), with edges between interacting atoms (vertices) representing covalent and non-covalent bonds. Different types of constraint graphs, such as the bar-joint,[19,20] the body-bar,[21,22] or the equivalent body-bar-hinge graph[23,24] all share the concept of assigning a number of pebbles to atomic vertices as DoF, and a number of bars to each interaction as constraints. Typically, rotatable single-covalent bonds and hydrogen bonds retain a dihedral degree of freedom, while peptide and double covalent bonds are modeled rigid. More constraints, such as additional hydrogen bonds, increasingly rigidify the graph. The remaining flexibility, i.e., the coordinated motion of internal DoF that do not perturb constraints, also known as *floppy modes*,[19,25,26] can be determined via constraint counting through the pebble game algorithm.[3,19] Constraint counting on these graphs has been applied to examine conformational flexibility in macromolecules,[20,27–30] probe the effects of ligand binding,[31] or estimate thermo-stability[32–34] and entropic measures.[35–37] Rigidity informa-tion has also helped to shape successful perturbation strategies in complex motion planning algorithms.[38–40] However, the topological nature of this approach denies explicit access to the kinematics of the underlying 3-dimensional molecular structure. Importantly, the exact motion vectors determined by the constraints (covalent and non-covalent bonds) remain unknown and can only be approximated by randomized perturbations[39] and iterative loop-closure algorithms such as ROCK[41] or FRODA.[42] While previous efforts tried to combine pebble game rigidity analysis with ENM,[43,44] these impediments have so far prevented a detailed comparison between the two methods.

Here, we overcome these limitations by using a kinematic approach to characterize flexibility and rigidity. Our approach explicitly provides basis vectors for motions corresponding to floppy modes, allowing us to analyze and compare motions directly from ENM and rigidity analysis. We previously established that topological rigidity and kinematic analysis yield an identical decomposition of the protein into rigid clusters for constraint compliant motions in non-singular configurations.[45] Topological rigidity analysis fails for protein conformations corresponding to singular kinematic configurations, where kinematic analysis gives correct decompositions.[45] Furthermore, here we exploit that our kinematic analysis can provide a basis for motions that progressively perturb constraints, i.e., motion modes that are inaccessible to topological rigidity analysis. In the remainder, we distinguish between *topological rigidity* as counting on constraint graphs via the pebble game, and *kinematic flexibility* as our new method.

Our study makes three main contributions. First, we establish that the hydrogen bonding pattern of a protein structure, together with its rotatable bonds, encodes a hierarchy of proteins motions. These orthogonal motion 'modes' are ranked by perturbations of the hydrogen bond network energy. Second, we show that this motion spectrum is conserved across a large dataset of different protein structures, but simultaneously reveals fold-specific differences. This fold-specific, spatial hierarchy suggests that hydrogen bond networks and folds are designed to modulate protein structure *and* dynamics. Third, we derive a formal, qualitative expression for a mode-specific free energy, which displays a near constant regime where increasing energetic cost is compensated by elevated entropy. Thus, our kinematic analysis provides a conceptual model system for enthalpy-entropy-compensation, with a readily accessible interpretation of the controversial phenomenon. Collectively, our results signify a deep connection between the structure of the hydrogen bond network, protein conformational dynamics, and functional motions. Our kinematic flexibility analysis is implemented in our KGS software, and available from https://github.com/ExcitedStates.

# Kinematic flexibility analysis of proteins

We model macromolecules as a kinematic linkage, with groups of atoms as rigid body vertices and rotatable covalent bonds as directed edges with a single degree of freedom (Figure 1, center). The linear, branched topology of monomeric molecules can be represented with a single kinematic spanning tree, rooted at an arbitrarily selected vertex (pink). Covalent double bonds, peptide bonds, and dihedrals amenable to planarity are modeled as rigid. Proline and aromatic rings are also modeled rigid, identical to standard topological approaches.[20,24] Groups of atoms without internal DoF are joined into rigid bodies. Dihedral angles of remaining single-covalent bonds are the DoF of the molecule, connecting two neighboring vertices. In contrast to topological rigidity, we distinguish between covalent DoF and non-covalent constraints such as hydrogen bonds. Without constraints (red, Figure 1 center), our model corresponds to a serial, open kinematic chain. Non-covalent interactions form closed kinematic cycles that coordinate dihedral motion, reducing the number of independent DoF. Multi-chain proteins can be treated within the same framework by connect-ing chains through inter-chain constraints or to a virtual super-root.[46,47] We encode hydrogen bonds as pentavalent holonomic constraints that allow only a rotation about the bond axis but prevent all other relative motion. In other words, the distance between the hydrogen and acceptor atom as well as the donor-hydrogen-acceptor and hydrogen-acceptor-base angles are constrained.[45] This corresponds to a constraint with five bars in pebble game algorithms, but explicitly specifies its kinematics. Other non-covalent interactions such as hydrophobic forces are not considered in this work. For a molecule with overall $n$ dihedral DoF, we can distinguish between free DoF $\mathbf{q}_f \in \mathbb{T}^f$ that do not appear in closed kinematic cycles, and constrained, cycle DoF $\mathbf{q} \in \mathbb{T}^d$, where $f+d=n$. The $m$ hydrogen bonds define a constraint variety $\mathcal{Q}$ on the cycle DoF

$$\mathcal{Q} = \{\mathbf{q} \in \mathbb{T}^d \mid \mathbf{\Phi}(\mathbf{q}) = \mathbf{0} \in \mathbb{R}^{5m}\}. \quad (1)$$

Differentiating the constraints with respect to time leads to instantaneous (velocity) constraint equations

$$\frac{d\mathbf{\Phi}}{dt} = \mathbf{J}\dot{\mathbf{q}} \begin{cases} = \mathbf{0}, & \text{if } \dot{\mathbf{q}} \text{ constraint observing} \\ \neq \mathbf{0}, & \text{if } \dot{\mathbf{q}} \text{ constraint violating} \end{cases}$$
(2)

which characterizes two disjoint, orthogonal subspaces corresponding to velocities that observe constraints and those that violate constraints. Here, $\mathbf{J} \in \mathbb{R}^{5m \times d}$ is the Jacobian matrix of the constraints. Constraint observing velocities span a subspace null $(\mathbf{J}(\mathbf{q}))$ of dimension $d-r$, with $r$ the rank of $\mathbf{J}$. Note that $r \leq p = \min(d, 5m)$ due to the rectangular shape of $\mathbf{J}$. The singular value decomposition (SVD,[48])

$$\mathbf{JV} = \mathbf{U\Sigma}, \quad (3)$$

where $\mathbf{U} = [\mathbf{u}_1, \ldots, \mathbf{u}_{5m}] \in \mathbb{R}^{5m \times 5m}$ and $\mathbf{V}[\mathbf{v}_1, \ldots, \mathbf{v}_d] = \in \mathbb{R}^{d \times d}$, provides orthonormal bases for the range and nullspace of $\mathbf{J}$. The rectangular matrix $\mathbf{\Sigma} = \text{diag}(\sigma_1, \ldots, \sigma_p) \in \mathbb{R}^{5m \times d}$ contains the singular values $\boldsymbol{\sigma}$ on the diagonal, where $\sigma_1 \geq \ldots \geq \sigma_r > \sigma_{r+1} = \ldots = \sigma_p = 0$, and corresponding $\mathbf{u}_i$ and $\mathbf{v}_i$ are the $i$th left and right singular vector, respectively. Let

$$\text{ran}(\mathbf{J}^T) = \text{span}\{\mathbf{v}_1, \ldots, \mathbf{v}_r\} =: \mathbf{R},$$
$$\text{null}(\mathbf{J}) = \text{span}\{\mathbf{v}_{r+1}, \ldots, \mathbf{v}_d\} =: \mathbf{N}. \quad (4)$$

Then, any $\dot{\mathbf{q}} = [\mathbf{R}, \mathbf{N}] [\nu_R^T, \nu_N^T]^T$ can be expressed with $\nu_i$ the proportion of velocity pointing along vector $\mathbf{v}_i$, respectively.

Since $\mathbf{u}_i$ and $\mathbf{v}_i$ are orthonormal vectors, $\sigma_i$ encodes non-orthogonality between $\mathbf{J}$ and $\mathbf{V}$; the singular value encodes the norm of constraint perturbation when moving along $\mathbf{v}_i$. Therefore, the two subspaces provide physically distinct insights. For motions along the nullspace, the perturbation is zero, encoded by the vanishing singular value. Thus, null($\mathbf{J}$) spans all constraint-observing motions. As shown previously, these modes are identical to floppy modes from topological rigidity approaches,[45] i.e., they are the only remaining internal DoF if no constraint perturbation is admitted. Dihedrals not part of these floppy modes are rigidified and merge adjacent rigid bodies, leading to a rigid cluster decomposition of the molecule, which is a central result from topological rigidity analysis. The nullspace is highly sensitive to the set of non-covalent constraints and careful tuning of energy cutoffs, e.g., for the inclusion of hydrogen bonds in protein and RNA, is crucial to prevent over-rigidification.[30] It is also noteworthy that the nullspace dimension scales nearly linearly with the size of the protein, i.e., the number of floppy modes increases with increasing dimensions of the Jacobian (Figure 2A).

The range ran($\mathbf{J}^T$) provides a spectrum of motions ranked by increasing constraint perturbation. Consistent with the notion in penalty methods,[49,50] the singular value $\sigma_i \geq 0$ denotes a penalty associated with cumulatively perturbing constraints when moving along singular vector $\mathbf{v}_i, i = 1, \ldots, r$. We term those motions *kinematic flexibility modes*. They are not accessible in topological rigidity analysis, since they do not respect constraints, but can potentially inform on hydrogen bonds that need to adapt their configuration during functional rearrangements or exchange with solvent. Thus, the full range of protein conformational dynamics encoded by the hydrogen bonding pattern is accessible through orthonormal bases spanning both matrix subspaces. We therefore refer to the SVD (3) of the constraint Jacobian obtained from the hydrogen bonding pattern as the *hydrogen bond network spectral decomposition*.

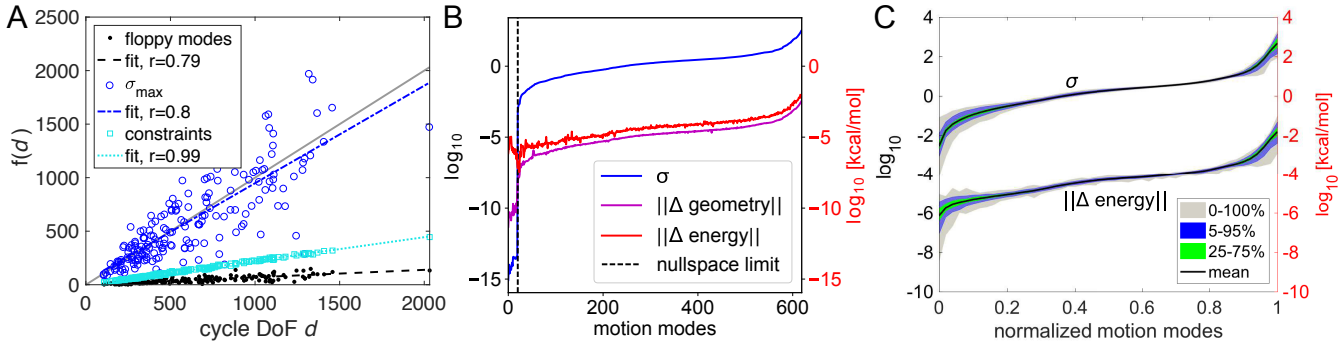

Figure 2: The hydrogen bond network spectrum $\boldsymbol{\sigma}$ predicts kinematic and energetic hydrogen bond perturbations, shown for an individual example (A, PDB ID 1p5f) with $d = 620$ and a large, diverse dataset with 183 high-resolution single-chain proteins (B). Modes are plotted in reverse index order, leading to a monotonous increase in $\boldsymbol{\sigma}$. Geometry and energy changes are computed from small perturbations (step size $\delta = 1e-5$) along kinematic flexibility modes. The ranking of energy perturbations by kinematic flexibility modes and the near log-linear motion regime are remarkably conserved among all proteins. Motion modes (horizontal axis) in panel B are limited to $\mathbf{R}$ and indices are normalized by $r$ to remove size-related differences in the dataset. (C) The number of hydrogen bond constraints, remaining floppy modes, as well as the magnitude of the largest singular value vary fairly linearly (correlation coefficient $r$) with the number of DoF $d$ in kinematic cycles, which serves as indicator for protein size.

## Results

### Kinematic flexibility modes hierarchically rank collective hydrogen bond energy perturbation

First, we analyzed how the hydrogen bond network spectral decomposition imposes a range of collective motions on the protein. To establish a predictive relationship between singular values and the *cumulative* variation in protein hydrogen bond geometry and energy, we monitored geometry and energy changes while taking a small step $\delta$ along the direction of each individual singular vector

$$\boldsymbol{\Delta}_{\mathbf{q},i} = \delta \mathbf{v}_i. \quad (5)$$

The vectors $\mathbf{v}_i$ are all unit length, and therefore $||\boldsymbol{\Delta}_{\mathbf{q},i}|| = \delta$, $i = 1, \ldots, d$. Changes in hydrogen bond geometry follow from $||\boldsymbol{\Phi}(\boldsymbol{\Delta}_{\mathbf{q},i})||$ (1), while energies $||\boldsymbol{\Delta E}_{\mathrm{HB}}(\boldsymbol{\Delta}_{\mathbf{q},i})||$ are evaluated using the Mayo energy potential[51] for each individual hydrogen bond

$$E_{\mathrm{HB}} = D_0 \left\{ 5 \left(\frac{R_0}{R}\right)^{12} - 6 \left(\frac{R_0}{R}\right)^{10} \right\} f(\theta, \psi, \phi), \quad (6)$$

with well-depth $D_0$, equilibrium distance $R_0$, hydrogen bond donor-acceptor distance $R$, and angular terms $f$ that depend on the hybridization state of donor and acceptor (details see SI (11)). We illustrate our findings using the crystal structure of 189-residue human DJ-1 protein (PDB ID 1p5f). Figure 2B shows monotonically increasing singular values (blue) which reversely rank-order the kinematic motion modes along the horizontal axis. The magnitudes of the corresponding cumulative geometric perturbations (magenta) and hydrogen bond energies (red) are shown for $\delta = 1e-5$.

The three curves clearly follow similar trends. Strikingly, while our kinematic flexibility anal-ysis is informed only by individual, geometric hydrogen bond information, cumulatively the hydrogen bond network spectral decomposition rank-orders protein motions hierarchically by geometry and energy penalty (Figure 2B). This suggests that the network is designed to selectively favor certain directions of collective motion in conformation space over other motions. Except for the smallest kinematic flexibility modes, near the nullspace, or the largest modes, the norm of the perturbations follow a log-linear regime. Modes outside the log-linear regime may correspond to non-functional protein dynamics. For example, modes with the largest perturbations correspond to unfolding motions where tertiary and secondary elements lose structure. Modes in the nullspace, i.e., floppy modes, carry no geometric penalty.

Next, we examined if the change in internal (hydrogen) bond energy could be predicted from the spectral decomposition. Note that the norm of constraint perturbation on velocity level and the singular values $\sigma_i$ are related by a scale factor $c_V = \delta = 1e-5$, i.e., $||\mathbf{J}\mathbf{\Delta}_{\mathbf{q},i}|| = c_V \sigma_i$ (combining (3) and (5)). Fitting the singular values curve (blue) to the magnitude of geometric perturbations of hydrogen bonds $||\mathbf{\Phi}(\mathbf{\Delta}_{\mathbf{q},i})|| = c_G \sigma_i$ (1), we found a scale factor of $c_G = 1e(-5.02 \pm 0.05) \approx \delta$, indicating that linearization has negligible effects at small step sizes. For the highly non-linear energy function ((6) and SI (11)), we found $c_E = 1e(-4.51 \pm 0.17) \approx 3.09\delta$. Hence, for a sufficiently small step size $\delta$, the hydrogen bond network spectral decomposition predicts the cumulative hydrogen bond energy perturbation along mode $i$ by $||\mathbf{\Delta E}_{HB}(\mathbf{\Delta}_{\mathbf{q},i})|| = c_E \sigma_i$ for this example structure. For floppy modes inside the nullspace, the difference between singular value prediction and energetic cost is larger. This can be explained by $sp2 - sp2$ hybridized hydrogen bonds that observe small energetic changes due to a torsional term in the hydrogen bond energy function (SI), which is not present in the geometric constraint formulation. The torsional term becomes more relevant if other terms such as distance and angles remain small, which is the case inside the nullspace.

This hierarchy of protein motions is remarkably conserved in the protein universe. We compiled a diverse benchmark dataset of 183 high-resolution, non-redundant crystal structures ranging in size from 30 to 555 amino acids in length from the PDB. We first observed an expected, strong linear relationship between the number of cycle DoF $d$ and the number of constraints $5m$ in the structures (Figure 2A). Table 1 evaluates statistical properties of the dataset. Similar to our single example crystal structure of human DJ-1 protein, singular values spanned many orders of magnitude. We also observed a large range in the number and distribution of vanishing singular values across the crystal structures, a median of 34 nullspace floppy modes, with a maximum of 148 and one (engineered) protein with zero floppy modes (PDB ID 5eca). While most structures have slightly more constraints than cycle DoF, i.e., $5m > d$, it appears that a fairly constant fraction of constraints is linearly independent, leading to a linear increase in the number of floppy modes over protein size (Figure 2A). This was also observed previously.[36] We therefore analyzed kinematic flexibility modes corresponding to non-vanishing singular values, i.e., outside of the nullspace, for all 183 crystal structures (Figure 2C). For comparison we normalized modes, and we grouped modes into 50 bins per structure. Repeating our analysis above, we obtained a surprisingly universal law from fitting the mean curves

$$(\mathbf{E}_{HB}^{\mathbf{q}})' = \frac{1}{\delta} \begin{pmatrix} \cdots \\ ||\mathbf{\Delta E}_{HB}(\mathbf{\Delta}_{\mathbf{q},i})|| \\ \cdots \end{pmatrix} = \bar{c}_E \boldsymbol{\sigma} \quad (7)$$

with $\bar{c}_E \approx 3.24$, *independent* of mode number, and error of the same order as before. This suggests that protein hydrogen bonding patterns impart a distribution of orthogonal, coordinated motions on the DoF. Kinematic flexibility modes identify preferred directions of deformation for the protein in (hydrogen bond) internal energy landscapes. Note that $(\mathbf{E}_{HB}^{\mathbf{q}})'$ reports on the cumulative energy change; motions along kinematic flexibility modes $i$ do not necessarily increase all hydrogen bond energies.

Table 1: Statistics of the high-resolution dataset with 183 single-chain proteins.

|        | # of residues | $m$ | $d$  | $d-r$ | h-bond energy   | $\sigma_1$ |
|--------|---------------|-----|------|-------|-----------------|------------|
| median | 163           | 116 | 516  | 34    | -422.94 kcal/mol  | 485.67     |
| min    | 30            | 23  | 107  | 0     | -1506.81 kcal/mol | 82.94      |
| max    | 555           | 445 | 2032 | 148   | -59.88 kcal/mol   | 1966.29    |

Instead, for each direction, fluctuations in hydrogen bond energies can provide compensatory mechanisms, i.e., many could marginally reduce in energy to allow a handful to significantly increase. Initial constrained minimization of the dataset ensures starting structures near a local minimum of hydrogen bond energy.

## Kinematic flexibility modes of $\alpha$-helices and $\beta$-sheets

To visualize motions across the spectrum and their effect on hydrogen bonds, we analyzed an $\alpha$-helix and a $\beta$-sheet in detail. The top panels of Figure 3 show the matrix product $\mathbf{JV}$, accumulated over the five constraints per hydrogen bonds, thereby graphically displaying equation (3). The predicted perturbation of individual hydrogen bonds (rows) for motions along each kinematic flexibility mode (columns) is color-coded, increasing from dark to light. For the $\alpha$-helix (A), we observed a striking, alternating pattern of perturbation, suggesting motions are modulated by concerted hydrogen bond energy fluctuations. We observed a similar, but weaker pattern for the $\beta$-sheet (B).

We then projected the conformational change $\Delta_{q,j}$ for degree of freedom $j$ corresponding to selected kinematic flexibility modes onto the helix and sheet (Figure 3, bottom panels). The left-most, purple-colored set of modes in the matrix correspond to local fluctuations (e.g., A11, B5, indices label column numbers), affecting only DoF close to individual hydrogen bonds shown as dots. As the mode number increases, DoF are increasingly engaged. For example, in super-helical twisting (e.g., A38) and straight bending (e.g., A46) compensating motions of DoF, indicated by alternating blue-red patterns, mediate moderate constraint perturbation. By contrast, high constraint perturbations found in compression/tension (A54/A55) are a result of changes in DoF that reinforce each other. Interestingly, unraveling (e.g., A19) demonstrates significantly less overall perturbation than compression/tension (A54/A55), indicating an energetically favorable mode for helix dissociation. Similarly, the structure of $\beta$-sheets permits twisting and bending motions (e.g., B11), while shear (e.g., B33) and especially widening/narrowing (e.g., B38) requires substantial distortion of hydrogen bonds. Our findings agree well with molecular dynamics simulations on the mechanical response of typical protein building blocks.[52] They revealed high strength of $\beta$-sheet proteins in response to shear loading, with an elastic modulus of about 240pN/Å. Also, the simulation results predicted an initial linear elastic regime during tensile loading of $\alpha$-helical protein domains. Beyond the elastic regime the helix uncoils, releasing one turn at a time. It is worth noting that there are no nullspace floppy modes along the backbone in either structure, i.e., topological rigidity analysis predicts only side-chain flexibility, but full backbone rigidity.

Besides the graphical interpretation, these results hint at two important characteristics of the motion spectrum. First, the perturbation patterns confirm distinct motion regimes, separated by perturbation levels and collectivity, i.e., how much of the structure is engaged in each mode. Second, the hydrogen bonding networks encode fold-specific motions. We examine these attributes quantitatively in the following.

## Spectral distribution of the hydrogen bond network

To obtain a physical understanding of the distribution of modes across protein structure, we performed a full spectral decomposition on our benchmark dataset. The size distribution of protein structures is reflected in the SVD of associated constraint Jacobian matrices $\mathbf{J}$. It is well-known mathematically that the maximum singular values monotonically increase by adding a column (a degree of freedom), while the smallest non-zero singular values decrease.[48] Physically, a larger protein with more DoF can be interpreted by a larger lever arm that allows a larger maximum constraint perturbation, or simultaneously provides more motion capabilities to avoid perturbation. Adding a row (constraint) will increase the minimum singular value, i.e., there are more constraints to be perturbed, and the protein is rigidified. Overall, the spectral range becomes size-dependent, as can be seen in the widened distributions near the spectral limits in Figure 2C. Hence, to compare mode densities across differently sized proteins, we normalized singular values of each structure by their maximum $\sigma_1$ to obtain $\sigma \in [0, 1]$ for all modes across all structures and grouped them by $\sigma$ into ten bins per decimal power.

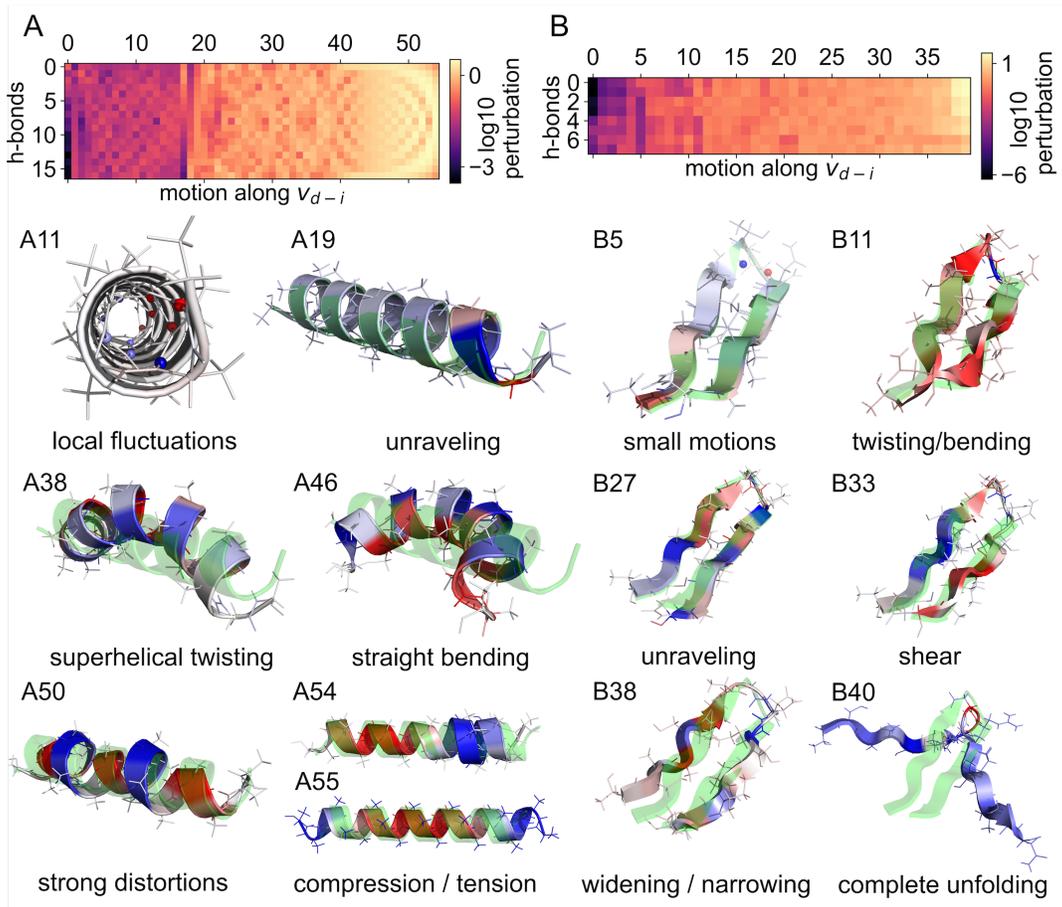

Figure 3: Hierarchical constraint perturbation in an $\alpha$-helix (A panels) and an anti-parallel $\beta$-sheet (B panels). The two top panels plot the matrix $\mathbf{JV}$, ranking singular vectors (columns) across individual hydrogen bonds (rows) by cumulative constraint perturbation, increasing from purple to yellow. Bottom panels depict motions along selected kinematic flexibility modes (column indices given). The motion amplitude is exaggerated (step size $\delta = 1$) for visualization purposes. Color coding indicates increasing change in DoF, from blue to red. The original conformation is shown in green.

Singular values for the 183 single-chain proteins span several orders of magnitude and show well-conserved, sharp peaks near integer exponents (Figure 4A). The two most common kinematic flexibility modes occur at $\sigma \approx 1e-2$ and $\sigma \approx 1e-3$. The peak at lowest $\sigma$ values (Figure 4A, far left) represents floppy modes, which are the only modes considered in topological rigidity analysis. Their density is least conserved across the spectrum and shows a linear increase over protein size (Figure 2A). Thus, while normalizing by $\sigma_1$ helps spectral comparison in the regime $\sigma > 0$, the variability in the number of floppy modes where $\sigma = 0$ remains.

The spectral distribution relates directly to the stiffness of the protein. Formally, this follows from defining a cumulative perturbation

$$p_c = \int_\sigma p(\overline{\sigma})\overline{\sigma}\mathrm{d}\overline{\sigma} \qquad (8)$$

for perturbation-specific probability densities $p(\sigma)$ following the spectral distribution, assuming individual modes are enabled at equal probability. Thus, proteins enriched in high-perturbation modes require overall more energy to access their motion modes, rendering them stiffer. Discrete jumps between spectral peaks suggest that modes are distributed across different stiffness regimes. Interestingly, atomic force microscopy (AFM) on single antibody proteins also found two distinct elastic regions with a ∼4-fold increase in stiffness between a low-strain and high-strain regime, before plastic deformation sets in.[53] Again, these experimental findings agree well with our perturbation analysis.

The geometry of hydrogen bonds in our analysis critically depends on accurately placed hydrogen atoms in structure preparation. To exclude the possibility of bias at that stage we repeated our analysis using a smaller set of 34 structures from the PDB with experimentally determined hydrogen atom positions from neutron diffraction experiments (SI). Its spectral signature, i.e., peak locations and heights, was virtually identical, validating the results from our initial dataset of 183 high-resolution single-chain structures for subsequent analysis. Furthermore, spectral analysis of singular values from a set of random matrices with the same dimensions as the original protein dataset produced a completely different distribution, while a set of random matrices with the same sparsity pattern led to similar distributions. This demonstrates that the hydrogen bonding network, together with the kinematic structure of the protein, stores this fold-specific dynamic information.

## Motions from the hydrogen bonding pattern are spatially distributed

Besides distinct perturbation levels, we observed various levels of mode engagement, i.e., how much of the molecule is involved in a specific motion mode. To measure this collectivity $s$ of mode $\mathbf{v}_i$, we compute the exponential of the Shannon entropy of its squared components[14,54]

$$s_i = \frac{1}{d}\exp\left(-\sum_{j=1}^{d}\kappa_{i,j}\log(\kappa_{i,j})\right), \qquad (9)$$

where $\kappa_{i,j} = \mathrm{v}_{i,j}^2/\sum_{j=1}^{d}\mathrm{v}_{i,j}^2$. This normalization of $\mathrm{v}_{i,j}$ is trivial for singular vectors, since they are unit length by definition, but non-trivial for eigenmodes from ENM or NMA. The second normalization by the number of modes $d$ (equals the length of $\mathbf{v}$) reduces size differences across protein structures, which allows $s_i \in [0,1]$ to be interpreted as the fraction of significant contributors in motion mode $i$.

Figure 4B plots collectivity computed over modes with similar singular values, grouped into ten bins per decimal power. Nullspace floppy modes with zero perturbation at the lower end of the spectrum show a relatively low collectivity compared to medium or high-perturbation modes. From panels A and B together we observe that the most collective motion modes are also most abundant. Collectivity, and thus entropy, follows a near-exponential regime between $1e-7$ and $1e-3$, showing a similar trend as energetic perturbations in Figure 2B. Finally, modes with strongest constraint perturbation are less collective than medium-perturbation modes.

## Protein fold classes have a unique h-bond network spectral signature

To analyze protein-fold specific differences encoded by the hydrogen bond pattern, we examined the hydrogen bond spectrum and col-lectivity of kinematic flexibility modes for four separate datasets of $\alpha$-only, $\beta$-only, $\alpha+\beta$ and $\alpha/\beta$ proteins, ranging from 655 to 1051 structures (SI). Figure 4 shows the mean curves for spectrum (C) and collectivity (D), corresponding to the black mean curves of the original mixed-fold dataset in (A) and (B), respectively. Percentiles follow similar patterns as for the previous data-set and are omitted for visibility. While the location of spectral peaks is conserved across folds, the peak heights show clear differences. The class of $\alpha$-only folds has more modes at higher singular values, yielding them stiffer to perturbations (8). By contrast, $\beta$-only proteins are enriched in smaller spectral modes, rendering them more flexible, while mixed folds $\alpha+\beta$ and $\alpha/\beta$ show intermediate spectra. The density of zero-perturbation floppy modes with vanishing singular values is highly size-dependent and less informative regarding fold content. Nonetheless, our results predict more floppy modes in $\alpha$-helical folds than $\beta$-folds, which can most likely be attributed to less inter-helical connectivity compared to inter-$\beta$-strand connectivity.

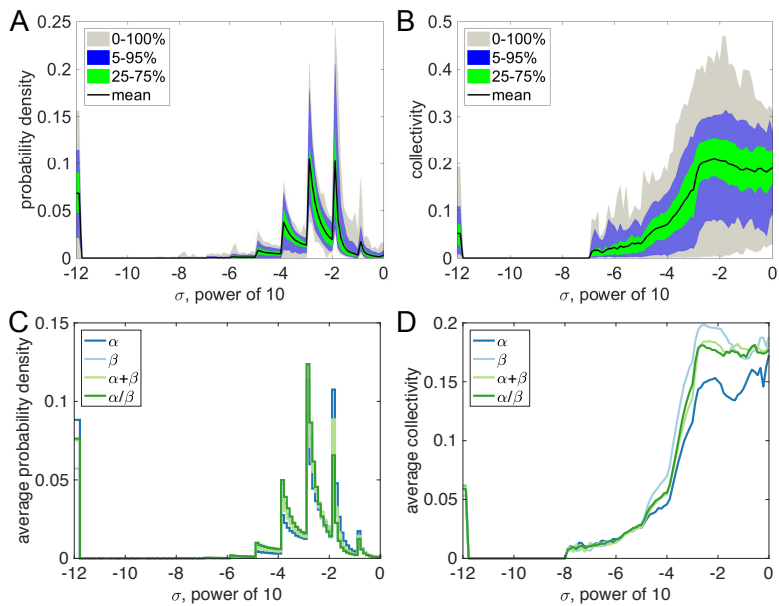

Figure 4: Kinematic flexibility analysis with hydrogen bond constraints reveals fold-specific motions in proteins. (A) Spectrum of normalized singular values for 183 single-chain proteins, with striking, well-conserved peaks. The nullspace dimension (peak at lowest $\sigma$) varies with protein size. (B) Collectivity of motion from Shannon entropy. Floppy modes with zero perturbation (lowest $\sigma$) show relatively small collectivity compared to modes at higher perturbations. Across a medium-perturbation range, collectivity increases with increasing singular values. (C) Fold-specific, average spectrum for classes of $\alpha$-only, $\beta$-only, $\alpha+\beta$ and $\alpha/\beta$ proteins (stair representation only for better visibility). Peak locations are well conserved across folds, while peak heights are shifted to lower $\sigma$ from $\alpha$-only to $\beta$-only. Floppy modes (lowest $\sigma$) vary rather with protein size than fold. (D) Average collectivity as in (B) for fold-specific datasets. Again, floppy modes with zero perturbation show relatively small collectivity; they are indistinguishable across folds.

Collectivity, computed as before, is generally higher for $\beta$-folds than $\alpha$-folds, and intermediate for mixed folds. Interestingly, medium-perturbation modes in $\beta$-dominated folds are more collective than the highest-perturbation modes, indicating localized hinges, e.g. in $\beta$-turns, that are able to significantly unfold the protein (cf. Figure 3B40). Note that floppy modes with zero-perturbation, the only accessible modes from topological rigidity analysis, show relatively small and very similar collectivity across folds, rendering them more localized compared to medium-perturbation modes and less informative regarding fold content.

To further evaluate predictive capabilities of our method, we analyzed the spectral distribution of a set of four hyperstable, designed peptides,[55] each with an NMR bundle of 20 distinct structures (PDB codes 2nd2, 2nd3, 5jhi, 5ji4; details in SI). Their spectrum shows a clear shift towards modes with higher relative perturbation compared to the high-resolution dataset (Figure 5) and a ∼3-fold increase in $p_c$ (8), identifying their designed constraint pattern as a key contributor to increased stability. Collectivity of modes shows trends as the other datasets.

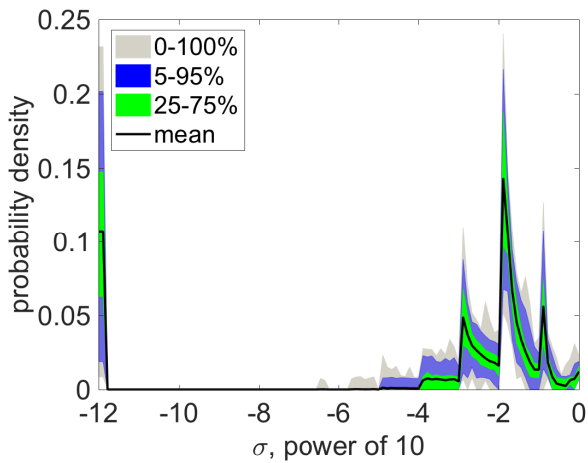

Figure 5: Spectrum of normalized singular values for four designed hyperstable, constrained peptides, each consisting of 20 NMR structures. The spectrum is shifted to higher perturbation modes relative to the high-resolution dataset.

## Free energy of modes

The hydrogen bond energy perturbations predicted by the magnitude of singular values can be formally combined with the conformational entropy contributions encoded by collectivity into a dimensionless expression for free-energy changes $\Delta F$ related to each mode $i$

$$\Delta F_i = \sigma_i - c_T s_i, \qquad (10)$$

with a dimensionless temperature factor $c_T$. For normalized singular values and normalized collectivities, the range of $\Delta F$ is between -1 and +1 when $c_T = 1$. Clearly, (10) is a formal abstraction, since only hydrogen bond energy contributes to internal energy. Surprisingly, though, our free-energy changes demonstrate how enthalpic and entropic contributions compensate each other in the overall spectrum of conformational motion. Figure 6 depicts $\Delta F$ over normalized motion modes computed from all 183 structures in the high-resolution dataset. The red curve averages over individual modes grouped into 100 bins. $\Delta F$ roughly levels for about 40% of modes (Figure 6 inset) with most favorable entropy (collectivity) and medium enthalpic cost. Modes to the right of the interval have unfavorably high enthalpic cost; they correspond to unfolding (high-perturbation modes in Figure 3). Modes to the left of the interval are more localized, with smaller enthalpic cost, but simultaneously less entropic benefit. Nullspace floppy modes are excluded in the graph; their enthalpic cost is zero, as encoded by vanishing singular values. Thus, associated free energy is identical to their collectivity as depicted in Figure 4B, which is on average $\Delta F = -0.05$ and thus turns out less favorable than the free energy plateau.

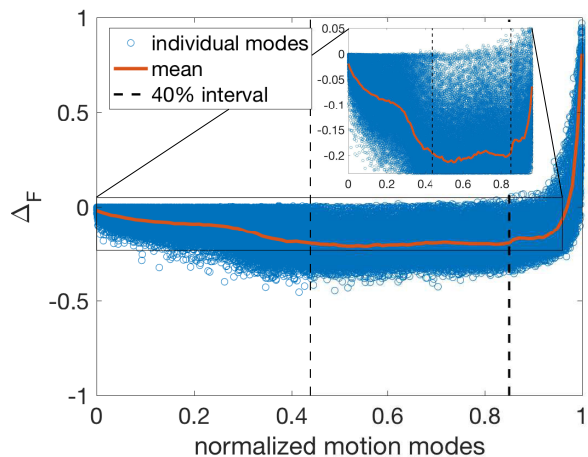

Figure 6: Dimensionless free energy of modes demonstrates entropy-enthalpy compensation encoded by the hydrogen bonding pattern. A 40% indicates a near constant free-energy level for highly collective modes at acceptable enthalpic cost (inset). Motions to the right correspond to unfolding, while motions to the left are more localized, with smaller entropic benefit.

## Comparison with iMOD

Finally, we compared the spectrum and collectivity of motions obtained from kinematic flexibility analysis with normal modes from iMOD.[4] iMOD is a versatile ENM-based tool to study normal modes of macromolecules in internal coordinates, i.e., dihedral angles. Although iMOD's mass matrix is based on a full atom representation, flexibility is limited to main-chain and $\chi_1$ angles, while our model maintains full side-chain flexibility. We carried out iMOD simulations with the command line settings $-n$ 10000 to force computation of all normal modes, $-x$ to enable the $\chi_1$ degree of freedom, and otherwise default parameters. Comparison of reduced ENM models to full NMA previously revealed that agreement of low-frequency modes is conserved, but that higher-frequency modes can differ significantly.[56]

Figure 7 shows the iMOD spectrum of eigenfrequencies (A) and collectivity of eigenmodes (B) for our 183 single-chain protein dataset. Similarly to our spectrum, eigenfrequency distributions are broadly conserved across structures, confirming results from previous NMA and ENM analysis.[11,56] Modes with low to medium eigenfrequency are most abundant. Similar to kinematic flexibility, the most abundant modes are also most collective. Higher-frequency motions with $\omega > 1000 \text{cm}^{-1}$ cannot be obtained with this coarse-grained model and require full NMA. Zero-perturbation floppy modes in topological rigidity and our kinematic flexibility analysis correspond to zero-frequency modes in ENM,[27] i.e., modes with vanishing energetic cost. However, zero-frequency modes in ENM are often considered an artifact, as the network breaks down into multiple independent ones. These modes are often avoided by increasing the number of weak interactions, for example by increasing cut-off distances.[13] Floppy motions from topological rigidity therefore cannot be directly accessed with ENM.

Zero-perturbation floppy modes from topological rigidity analysis are less collective, i.e., correspond to more local motions compared to low-frequency normal modes or medium kinematic flexibility modes (compare Figure 4B to Figure 7B). This signifies that topological rigidity theory based methods tend to overestimate molecular rigidity, and study conformational flexibility based on more local motions than ENM or NMA.

When we compared fold-specific spectra (Figure 7C) and collectivity (Figure 7D), we observed several important differences between the methods. For example, the location and height of the main spectral peak in iMOD is indistinguishable for $\alpha$-only and $\beta$-only, and shifted slightly to higher frequencies for mixed folds, while kinematic flexibility analysis shows stronger differences in peak heights across fold types. Interestingly, $\alpha$-folds show slightly increased density at very low frequencies in iMOD, predicting lower stiffness compared to other folds, in contrast to other established methods. Our kinematic flexibility analysis, detailed NMA,[10,56] and analysis of force-displacement curves from MD simulations[52,57] all predict that $\alpha$-folds are stiffer. However, iMOD predicts lower collectivity in $\alpha$-folds, similar to our approach.

A more detailed comparison between the methods, e.g., analyzing direct mode overlap, remains difficult, since iMOD is limited to main-chain and $\chi_1$ dihedral motion, while our kinematic model maintains full side-chain flexibility, thus demanding lower-dimensional projections. Other ENM or NMA tools are mostly formulated in cartesian coordinates or are unavailable for other users, making a comparison even more difficult. Nevertheless, our results reveal similarities and differences important for users of ENM, topological rigidity, and our new kinematic analysis.

## Discussion

Our new, kinematic approach to rigidity analysis treats hydrogen bonds as a geometric constraint network. Compared to topological rigidity, it extends analyses to constraint perturbing motions, providing a full spectral decomposition of motion modes ranked by their cumulative hydrogen bond energy perturbations. While analyses of the structural dynamics of proteins often implicitly assume that weak hydrogen bonds disrupt first, our approach does not require such assumptions. Instead, the network imparts a hierarchy of motions and stiffness regimes on the protein, which modulate the conformational response. The hydrogen bond spectral decomposition is 1) highly conserved in the protein universe, and 2) reveals key fold-specific differences.

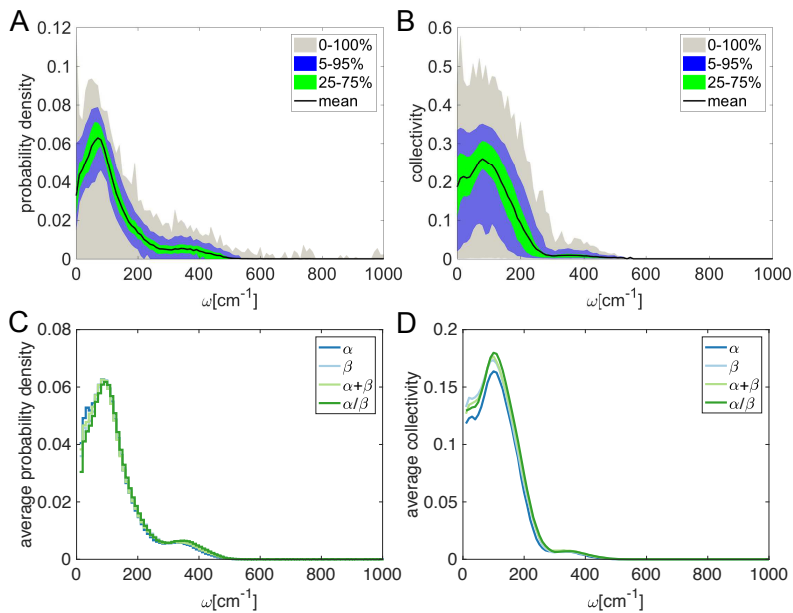

Figure 7: Eigenspectrum and collectivity for normal modes computed with iMOD. (A) Eigenfrequency spectra of the data-set with 183 single-chain proteins. (B) Collectivity of eigenmodes. (C) Eigenfrequency spectra of four fold-specific data-sets. (D) Fold-specific collectivity of eigenmodes.

Kinematic flexibility analysis indicated that zero-perturbation floppy modes, the motions obtained from topological rigidity analysis, are more localized than low-frequency modes from ENM, suggesting an overly rigidified representation in topological rigidity or alternatively, over-connectivity in ENM. Therefore, strict rigidity theory based methods rely on exceedingly local motions to estimate conformational flexibility,[20,27,28] ligand binding,[31] entropy,[36,37] or thermo-stability[32–34] compared to ENM or NMA. Nonetheless, they show convincing agreement with experimental data. Moreover, kinematic floppy modes that observe hydrogen bond constraints have guided conformational transitions[39,58,59] and revealed coordinated loop motions,[60] often more successful than normal mode based methods.[47,58]

Analyses of motions beyond the floppy modes, i.e., in the kinematic flexibility regime, revealed several unexpected insights. First, we found that hydrogen bond networks determine structure *and* modulate structural dynamics. This could have important implications for *de novo* protein design and folding[61] or hydrogels,[62] where recent attention is focused on designing hydrogen bonding patterns to create stable interfaces[63] and mediate specificity.[64,65] For example, our procedure revealed a clear shift toward stiffer modes for designed, hyperstable peptides.[55] While hydrogen-bond guided designs are often successful structurally, i.e., the crystal and predicted structure are near identical, it remains a challenge to design dynamic, functional proteins.[66] Our procedure could predict motion modes of hundreds of designed proteins and hydrogen bond networks in minutes. Second, the conserved hierarchy and collectivity of protein motions revealed distinct motion regimes, which are often observed in experiments. For example, AFM on single antibody proteins[53] uncovered two elastic motion regimes, separated by a near 4-fold increase in stiffness, and a regime of plastic deformation. Our spectral decomposition structurally rationalizes the distribution of these motions under different strains. Low strain triggers low-perturbation modes that are low in collectivity and mostly locally perturb the structure. Higher strain engage highly collective modes toward stiffer motion regimes, explaining elevated stiffness in the second elastic regime measured with AFM. Interestingly, our EEC model suggests that a fraction of the energetic cost could be entropically balanced, rendering the deformation elastic and reversible.[53,67] Any strain beyond the elastic regime leads to plastic deformation, indicated by a sharp increase in hydrogen bond energy perturbation. Motion modes in this regime likely completely unfold the protein.

Third, the fold-specific differences we observe in our perturbation analysis suggest distinct functional roles of secondary structure, confirming previous simulations.[52,57] We found more modes at higher energy perturbation in $\alpha$-helices than $\beta$-sheets, yielding helices stiffer. This is consistent with experimental data from low-frequency Raman spectroscopy[68] and eigenfrequencies from detailed NMA.[10,56] Detailed NMA captures additional differences between folds in high-frequency regimes such as the amide I, II, or III bands.[56] These higher-frequency differences can also be detected experimentally with infrared spectroscopy and have been used to determine secondary structure content.[69,70] Simplified elastic network analysis using iMOD failed to identify this shift toward stiffer modes, highlighting a limitation in the simplified force-fields of most ENM. Our analysis also predicted that $\beta$-sheets are highly resistant to shear motions, which is confirmed by MD simulations.[52] Simulations on the mechanical response of silk crystalline units under shear loading showed high rupture forces due to efficient force distribution in the $\beta$-sheet structures.[71] Buehler and Keten further found an initial linear elastic regime during tensile loading of $\alpha$-helical proteins, after which the helix unravels turn by turn.[52] Our analysis also predicted unraveling as the favored mode of helix dissociation.

Our method can provide quick insight into how rigid clusters, collective motions, and stiffnesses shift as constraints are added or removed. While we found convincing agreement with experiment and simulation, important limitations remain. Our model considers only intra-molecular hydrogen bonds, ignoring other non-covalent interactions. Bond lengths and angles are fixed and only torsional DoF contribute to molecular motion. While hydrogen bonds are important determinants of protein structure and dynamics, hydrophobic effects, electrostatics, solvent, or protein-protein interactions also modulate structural dynamics. The effects of these interactions could be explored with our method. For example, neutron diffraction reveals the position and orientation of hydrogen atoms in waters. Adding water-mediated hydrogen bonds can help understand how solvation propagates collective motions in protein cavities or binding pockets.

Our new, kinematic flexibility analysis is a versatile method, bridging topological rigidity and ENM. By way of a novel spectral decomposition of protein hydrogen bonding patterns, it provides explicit access to collective motions and free energy of modes, signifying that hydrogen bonds store intrinsic, fold-specific functional motions. These quantitative models and insights can help improve *de novo* protein design and folding, help to understand mechanobiology probed by AFM or single-molecule fluorescence resonance energy transfer (smFRET) at the molecular level, or help interpret experimental data such as hydrogen deuterium exchange.

**Acknowledgement** The authors gratefully acknowledge financial support to D.B. from the Deutsche Telekom Stiftung. H.v.d.B. is supported by NIH GM123159.

# Datasets

The initial high-resolution dataset of 183 proteins was obtained through an advanced entity-based PDB search limited to crystallographically collected single-chain protein structures between 30 and 1000 amino acids in length, with a single oligomeric state and model, at most 50% sequence identity and no free or modified residues ligands. To ensure high validity, we required a resolution of at least 1.5 $\text{\AA}$ as well as $R_\text{free}$ and $R_\text{work}$ values below 0.2. The dataset was then curated with Schrodinger's Maestro suite filling small gaps, performing a constrained minimization with an RMSD of 0.3 (0.4 in one single case where no minimum was found otherwise), and optimizing hydrogen positions for hydrogen bonding. One structure, 2qvk, had to be excluded due to a large missing domain. Water molecules were included in the Maestro preparation step, but ignored in subsequent KGS analysis.

The neutron scattering control dataset was similarly obtained from the PDB, specifying experimental method to neutron diffraction with experimental data present and limiting the search to single-chain proteins at 90% sequence identity, leaving 34 structures in the dataset. Structures were also prepared using Maestro, this time without resampling hydrogens to preserve experimentally determined positions and orientations.

The fold-specific datasets were obtained from the PDBs SCOP classification, limited to single-chain, single oligomeric state proteins with at most 90% sequence identity, leading to 655 $\alpha$-only, 877 $\beta$-only, 885 $\alpha+\beta$, and 1051 $\alpha/\beta$ examples, respectively. Due to the size of the datasets, we refrained from full Maestro-based preparation and added hydrogen atoms with Reduce instead.

The four constrained peptides were downloaded from the PDB via accession codes 2nd2, 2nd3, 5jhi, and 5ji4, respectively. Each of their twenty individual NMR structures was processed like the fold-specific datasets, by adding hydrogens with Reduce and successively identifying hydrogen bond constraints.

## Hydrogen bond identification and energy computation

Hydrogen bonds are traditionally identified based on stringent geometric criteria on hydrogen-acceptor distance, acceptor-donor distance, and various angles between them. We developed our own Python-based implementation using criteria developed in HBPlus,[72] limited to strong hydrogen bonds between an electro-negative donor and acceptor (oxygen, nitrogen, or sulfur).

There are several ways to estimate hydrogen bond energy.[72–74] Different approaches include electrostatic modeling,[75] quantum-mechanical modeling based on electron density,[76] and extensions with the density's Laplacian.[77] All these, however, require detailed experimental data and are thus quite complex.

Traditionally, the Mayo energy function is used in topological rigidity analysis[20] (see full text (6)). Its full definition[51] is given by

$$E_\text{HB} = D_0 \left\{ 5 \left(\frac{R_0}{R}\right)^{12} - 6 \left(\frac{R_0}{R}\right)^{10} \right\} f(\theta, \psi, \phi), \tag{11}$$

with $R_0 = 2.8\text{Å}$ and $D_0 = 8\text{kcal/mol}$ are the corresponding equilibrium distance and well-depth, and $R$ the donor-acceptor distance. Angular terms depend on hybridization, with the following specifications:

$$\begin{aligned} \text{sp}^3\text{donor - sp}^3\text{acceptor:} \quad & f = \cos^2\theta \cos^2(\psi - 109.5), \\ \text{sp}^3\text{donor - sp}^2\text{acceptor:} \quad & f = \cos^2\theta \cos^2\psi, \\ \text{sp}^2\text{donor - sp}^3\text{acceptor:} \quad & f = \cos^4\theta, \\ \text{sp}^2\text{donor - sp}^2\text{acceptor:} \quad & f = \cos^2\theta \cos^2(\max[\psi, \phi]), \end{aligned} \tag{12}$$

where $\theta$ describes the donor-hydrogen-acceptor angle, $\psi$ the hydrogen-acceptor-base angle, and $\phi$ the angle between normals of the planes defined by donor and acceptor bonds. PDB based evaluations[78] show that a heuristic-based energy function might produce more realistic energies, especially for hydrogen bonds outside the classical geometric configuration favored by the Mayo potential. Here, however, we are interested in energetic changes, where we expect only marginal differences between both approaches, such that the commonly used energy function in rigidity analysis suffices.

For all datasets, hydrogen bonds were identified with an in-house python implementation, using kinematic criteria as defined in[72] and an energy cut-off of $-1\text{kcal/mol}$ using the above "Mayo" potential.[51] All hydrogen bonds stronger than that were included as constraints in our kinematic model. Other non-covalent interactions were not considered.

### iMOD

iMOD is developed and maintained by the Chacon lab and performs normal mode analysis with an elastic network model in internal, torsional coordinates.[4] For our comparative analysis, we ran iMOD on each structure in our datasets, setting option $-n$ 10000 to force the maximum number of normal modes to be calculated. We specified option $-x$ to enable the $\chi_1$ dihedral degree of freedom and chose default options otherwise. This representation considers all heavy atoms individually, but only mainchain and $\chi_1$ flexibility in the resulting modes.